\documentclass{PoS}

\usepackage{graphicx}
\usepackage{pstricks}
\usepackage{amsmath}
\usepackage{amssymb}
\usepackage{bm}
\usepackage{booktabs}
\usepackage{psfrag,graphicx}
\usepackage{cite}
\usepackage{mathrsfs}
\usepackage{type1cm}
\newcommand{\Comments}[1]{}
\usepackage[normalem]{ulem}  % \sout{old text} for strikeout
\newcommand{\com}[1]{{\color[rgb]{0,0,1}{#1}}}
\newcommand{\comm}[1]{{\color[rgb]{0,1,0}{#1}}}
\newcommand{\comment}[1]{}
\renewcommand{\sout}{\bgroup \color{red} \ULdepth=-.5ex \ULset}
\renewcommand{\com}[1]{#1}
\renewcommand{\comm}[1]{}
\renewcommand{\sout}[1]{}
\newcommand{\tcom}[1]{{\color[rgb]{0,0,1}{#1}}}
\newcommand{\tcombar}{\bgroup \color{red} \ULdepth=-.5ex \ULset}
\renewcommand{\tcom}[1]{#1}
\renewcommand{\tcombar}[1]{}
\newcommand{\bk}{\mathbf{k}}
\newcommand{\mrm}[1]{\mathrm{#1}}
\newcommand{\expv}[1]{\left< #1 \right>}

\newcommand{\equref}[1]{Eq.~(\ref{#1})}
\newcommand{\figref}[1]{Fig.~\ref{#1}}
\newcommand{\arcsinh}{\mathrm{arcsinh}}
\newcommand{\vk}{{\bold{k}}}

\newcommand{\vkt}{{\vk,\tau}}

\title{QCD phase diagram at strong coupling including auxiliary field fluctuations\footnote{Report No. : 
  YITP-13-118, 
  KUNS-2469
}}

\ShortTitle{QCD phase diagram at strong coupling including auxiliary field fluctuations}

\author{\speaker{Terukazu Ichihara}\\%
        Yukawa Institute for Theoretical Physics \& Department of Physics, \\
        Kyoto University, Kyoto 606-8502, Japan\\
        E-mail: \email{t-ichi@ruby.scphys.kyoto-u.ac.jp}}

\author{Takashi Z.~Nakano\\
        Kozo Keikaku Engineering Inc.,\\
        Tokyo 164-0012, Japan\\
        E-mail: \email{takashi-nakano@kke.co.jp}}

\author{Akira Ohnishi\\
        Yukawa Institute for Theoretical Physics,\\
        Kyoto University, Kyoto 606-8502, Japan\\
        E-mail: \email{ohnishi@yukawa.kyoto-u.ac.jp}}

\abstract{
We investigate the \com{phase diagram and the mechanism for} 
the sign problem
\com{to appear}
\com{in finite density} QCD 
at strong coupling 
\com{in a combined framework of the} auxiliary field Monte-Carlo (AFMC) \com{ and the chiral angle fixing \tcom{(CAF) }methods}.
When bosonizing  meson hopping term\com{s in the effective action},
we need to introduce 
\com{imaginary number coefficients},
which leads to a complex phase in  numerical simulation\com{s}. 
\com{By using the cut-off technique, we} 
quantitatively show that high momentum modes of auxiliary fields
\com{mainly} contribute to the 
\com{weight cancellation;}
\com{Cutting-off high-momentum auxiliary field modes 
does not modify order parameters but suppresses statistical weight cancellation,
when we choose the cut-off parameter appropriately.}
}

\FullConference{31st International Symposium on Lattice Field Theory - LATTICE 2013\\
		July 29 - August 3, 2013\\
		Mainz, Germany}

\begin{document}

\section{Introduction}
QCD phase diagram is one of the challenging subjects 
\com{in} quark-hadron \com{sciences}.
Lattice QCD is one of the powerful tools to investigate the QCD \tcom{phase}
\com{transition} 
especially \com{in the} low chemical potential ($\mu$) region.
However, we confront a notorious sign problem at finite $\mu$.
Chemical potential makes the statistical weight complex.
\tcom{It} leads to the weight cancellation and lower reliability 
of the simulation especially \tcom{on} a large lattice. 
Therefore, it is difficult to simulate finite $\mu$ region.

Strong coupling lattice QCD (SC-LQCD) is 
\com{promising machinery} to attack 
the sign problem. 
SC-LQCD is based on 
\com{the expansion of the lattice QCD effective action in} 
inverse coupling ($1/g^{2}$). 
We \com{can }study the chiral phase transition~\cite{kawamotosmit}
and the QCD phase diagram 
\tcom{\cite{Bilic92-2,Phasediagram,NLOandNNLO}.}
Most of previous works are carried out under the mean field (MF) 
approximation\com{, where there 
is no sign problem}. 
\com{When we take account of fluctuations, 
however, we have sign problems as found} 
in both monomer-dimer-polymer (MDP) simulations~\cite{Fromm,Unger} 
and auxiliary field Monte-Carlo (AFMC) method~\cite{Ohnishi:2012yb}.

\begin{figure}[tb]
%\PSfig{7.5cm}{L4N4.eps}~\PSfig{7.5cm}{PB6.eps}
  \begin{minipage}{0.5\columnwidth}
    \begin{center}
      \includegraphics[width=0.58\columnwidth,angle=270]{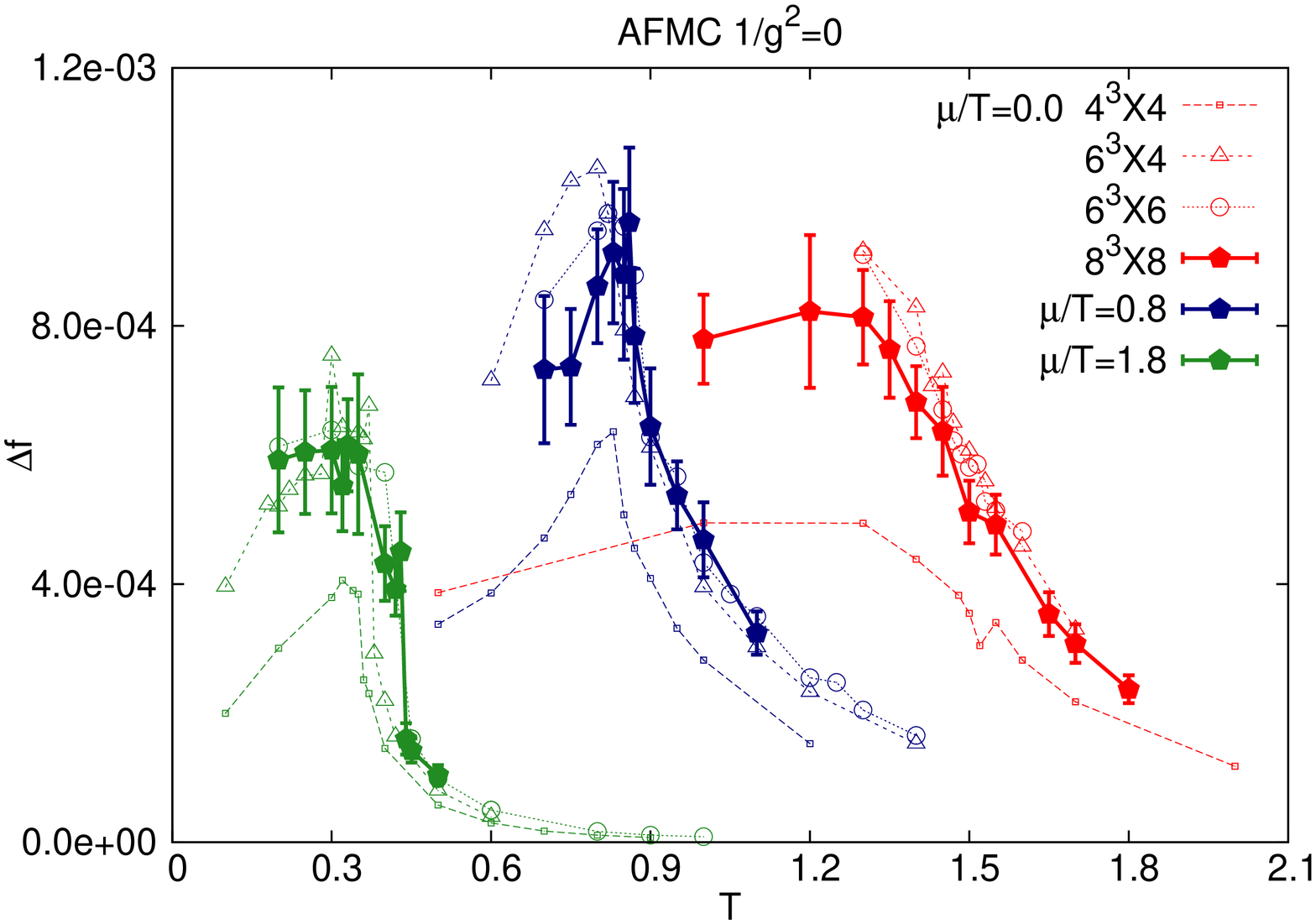}
    \end{center}
  \end{minipage}
%\hfill
%~\hspace*{1cm}~
  \begin{minipage}{0.5\columnwidth}
    \begin{center}
      \includegraphics[width=0.58\columnwidth,angle=270]{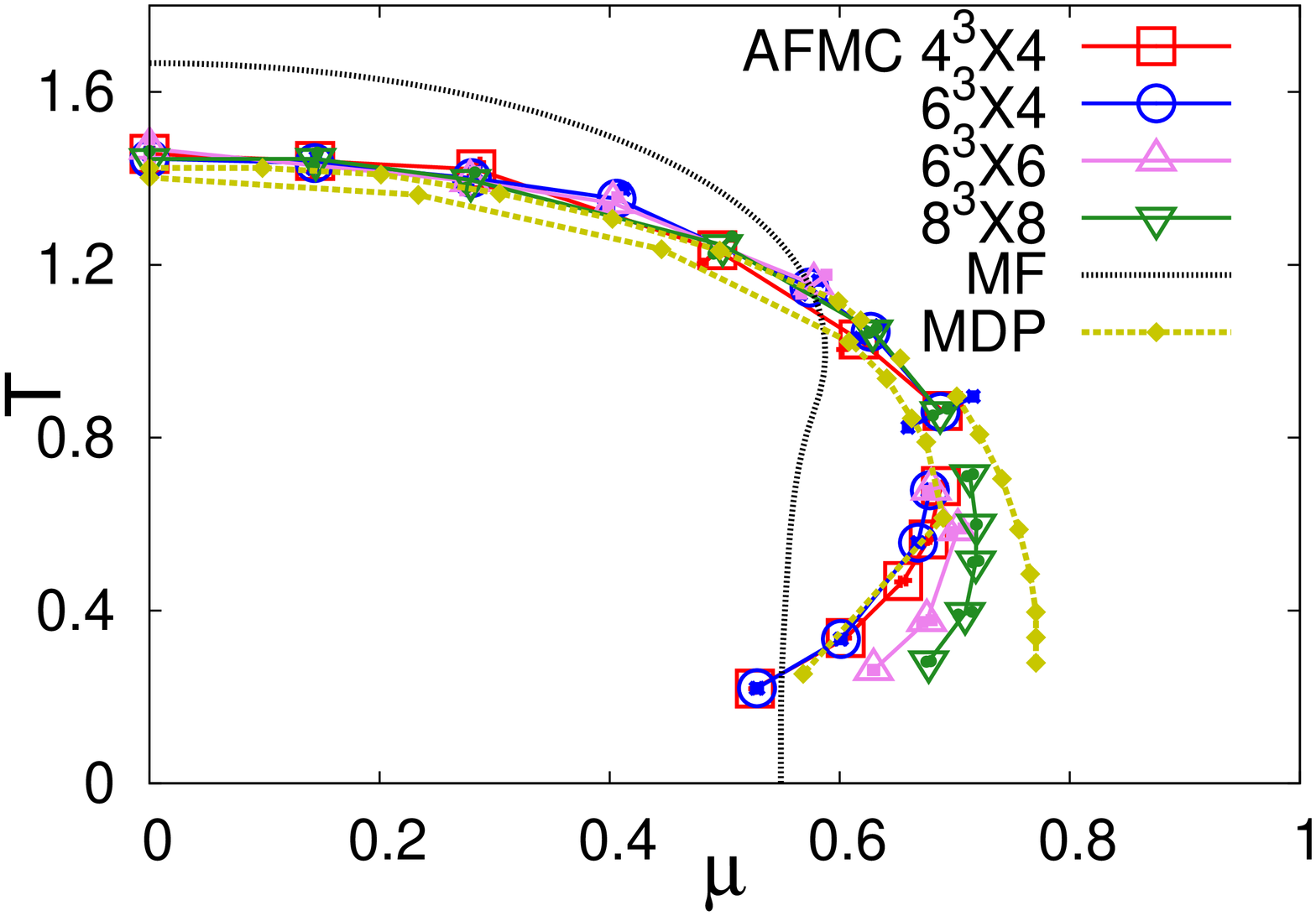}
    \end{center}
  \end{minipage}
\caption{Diff\tcom{e}rence of free energy density between full 
and phase quenched MC simulations, $\Delta f$, in AFMC 
as a function of temperature (left panel) and phase boundary (right panel). 
In left panel, we only show the jack-knife error bars in the results on a $8^{4}$ lattice. 
In right panel, we show various lattice size results in AFMC. MF denotes the mean field result. 
MDP is the result based on Monomer-Dimer-Polymer (MDP) simulations \cite{Fromm,Unger}.
}
\label{Fig:dfPB}
\end{figure}

It is important to know how sever\com{e} the sign problem is
in both MDP and AFMC to analyze the QCD phase
diagram on a large lattice. 
The severity of the sign problem is characterized by the difference
of the free energy density, $\Delta f = f_{\mrm{full}} - f_{\mrm{p.q.}}$,
in full Monte-Carlo (MC) and phase quenched MC simulations\tcom{;
$\Delta f$ is related to the average phase factor as 
$e^{- \Omega \Delta f} =\expv{e^{\mrm{i}\delta}}$, 
where $\Omega$ and $\delta$ are the space-time volume 
and the complex phase in each configuration, respectively.}
%%%%%%%%%%%%%%%%
\com{Comparison of} $\Delta f$ 
\com{in MDP~\cite{DrFromm} 
and AFMC (shown in \figref{Fig:dfPB})}
\com{shows} that $\Delta f$ of AFMC is 
twice as large as that of MDP.

In this paper, we \com{discuss the QCD phase
diagram and the mechanism for} the sign problem \com{to appear} 
in \com{AFMC, where} 
we integrate out auxiliary fields exactly by using MC technique.
We here \com{examine} the high momentum auxiliary mode contributions to the weight cancellation and order parameters. 
These analyses are \com{carried out 
in the}
chiral angle fixing (CAF) method\com{, where we fix the chiral angle
in each configuration and we can guess an appropriate chiral limit in a large
volume}. 
Finally, we give the QCD phase diagram obtained in AFMC with CAF. 

\section{Formalism} \label{Sec:formalism}

\subsection{Effective action \& AFMC method}
In present article, we consider the case where $SU(N_c = 3)$ 
in 3+1 dimension $(d = 3)$ space-time. 
\com{Temporal} and spatial lattice sizes 
\com{are} $N_\tau$ and $L$.
In the strong coupling limit, \com{we can ignore }the plaquette term in the lattice QCD action.
After integrating out spatial link variables in the leading order 
of the $1/d$ expansion, 
the strong coupling limit (SCL) effective action
\tcom{
\cite{kawamotosmit,Ohnishi:2012yb,Bilic92-2,Phasediagram,NLOandNNLO,faldt}
}
with one species of unrooted staggered fermion is given as,
\begin{align}
S_\mathrm{eff}
&=\frac12 \sum_x \left[ V^{+}_x - V^{-}_x \right]
- \displaystyle \frac {1}{4N_c} \sum_{x} M_x M_{x+\hat{j}}
+m_0 \sum_{x} M_x 
\ ,\label{Eq:Seff}\\
V^{+}_x&=\gamma e^{\mu/\gamma^2} \bar{\chi}_x U_{0,x} \chi_{x+\hat{0}}
\ ,\quad
V^{-}_x=\gamma e^{-\mu/\gamma^2} \bar{\chi}_{x+\hat{0}} U^\dagger_{0,x} \chi_x
\ ,\quad
M_x=\bar{\chi}_x \chi_x
\ ,
%\label{Eq:Seff}
\end{align}
where $\chi_x (\bar{\chi}_x)$ and $U_{\nu,x}$ denote the (anti-)quark field and the link variable. We introduce anisotropic
factor $\gamma$ as the ratio of temporal to spatial action
\com{coefficients}~\cite{Bilic92-2,Fromm,Unger,faldt,DrFromm},
and chemical potential $\mu$ as temporal component 
of vector potential. 
$\eta_{j,x}=(-1)^{x_0+\cdots+x_{j-1}}$ is the staggered sign factor.

Next, we bosonize 4-fermi like term\com{s} 
to integrate out Grassmann variables. 
Since mesonic fields ($M_x$) take different value at each site, 
we use the extended Hubbard-Stratonovich (EHS) 
transformation \cite{NLOandNNLO,Ohnishi:2012yb}.
\comment{
\begin{align}
e^{\alpha A B}  
= \int\, d\psi\, d\psi^*\,
	e^{-\alpha\left\{
	\psi^* \psi-{A}\psi-\psi^* B
	\right\}}
\ ,\ 
e^{-\alpha A B}  
= \int\, d\psi\, d\psi^*\,
	e^{-\alpha\left\{
	\psi^*\psi-i{A}\psi-i\psi^* B
	\right\}}\ ,
\label{Eq:EHSA}
\end{align}}

\begin{align}
e^{\alpha A B}  
= \int\, d\sigma\, d\pi\,
	e^{-\alpha\left\{
	\sigma^2+\pi^2
	+\sigma(A+B)+i\pi(A-B)
	\right\}}\ .
\label{Eq:EHS}
\end{align}
\equref{Eq:EHS} is an identical equation, so it is inevitable that we introduce an imaginary number coefficient. 
This is the origin of the sign problem in AFMC. 
After \com{spatial} Fourier
transformation ($M_{x=(\bm{x},\tau)}= \sum_{\bm{k}} e^{\mrm{i}\bm{k}\cdot\bm{x}}M_{\bm{k},\tau}$), we utilize the EHS transformation with respect to the eigen value of the composite mesonic fields (the second term of \equref{Eq:Seff}). 
Then, the auxiliary fields \com{for}
$M_{\bm{k},\tau}$ and $\mrm{i}M_{\bar{\bm{k}},\tau}$ 
\com{($\bar{\bm{k}}=\bm{k}+(\pi,\pi,\pi)$)} are introduced as $\sigma_{-\bm{k}}$ and $\pi_{\bm{k}}$~\cite{Ohnishi:2012yb}.
After bosonization, the action becomes
\begin{align}
S_\mathrm{eff}^\mathrm{EHS}
&=\frac{1}{2}\sum_x\left[V_x^+ - V_x^-\right]
 +\sum_x m_x M_x
 +\frac{L^3}{4N_c} \sum_{\vkt, f(\vk)>0}
f(\vk)\left[\left|\sigma_\vkt\right|^2+\left|\pi_\vkt\right|^2\right]
\label{Eq:SeffEHS}
\ ,
\\
m_x
&=
 m_0
 +\frac{1}{4N_c} \sum_{j}
	\left[
	 (\sigma+i\varepsilon\pi)_{x+\hat{j}}
	+(\sigma+i\varepsilon\pi)_{x-\hat{j}}
	\right]
\ ,\label{Eq:meff}
\end{align}
where $f(\bm{k})=\sum_{j=1}^{d} \cos k_j$,
$\sigma_x = \sum_{\vk,f(\vk)>0} e^{i\bold{k}\cdot\bold{x}}\sigma_\vkt$, and
$\pi_x = \sum_{\vk,f(\vk)>0} (-1)^\tau e^{i\bold{k}\cdot\bold{x}}\pi_\vkt$.
$\varepsilon_x=(-1)^{x_0+\tcom{\cdots}+x_3}$
plays a \com{similar} role \com{to} $\gamma_5$
in the continuum limit. 
We should notice that the imaginary number exists
in the spatial diagonal part\tcom{s} of the fermion matrix
from the modified mass terms.

Finally, we integrate out the Grassmann and 
temporal link ($U_0$) variables analytically \cite{faldt},
and obtain the effective action,
\begin{align}
S_\mathrm{eff}^\mathrm{AF}
=\sum_{\vkt, f(\vk)>0}
  \frac{L^3f(\vk)}{4N_c}
  \left[\left|\sigma_\vkt\right|^2+\left|\pi_\vkt\right|^2\right]%\ ,\nonumber \\
-\sum_{\bm{x}} \log\left[
	X_{N_\tau}(\bold{x})^3-2X_{N_\tau}(\bold{x})
	+2\cosh(\frac{3 N_\tau\mu}{\gamma^2})\right]
\ .
\label{Eq:SeffAF}
\end{align}
$X_{N_\tau}$ is a function of the modified 
mass ($m_x$)\com{, and obtained by using} a recursion 
formula\cite{faldt}.
In the MF approximation, we get $X_{N_{\tau}}=2 \cosh(N_{\tau} \arcsinh (m_x /\gamma))$.
\com{We} \com{numerically} carry out the auxiliary field ($\sigma_{\bk,\tau} , \pi_{\bk,\tau}$) integration 
\com{of the partition function, $Z=\int \mathcal{D}[\sigma,\pi] \exp(-S_\mathrm{eff}^\mathrm{AF})$, }
by using \com{the} Monte-Carlo method.

\subsection{Weight cancellation in AFMC}
\com{We have} the weight cancellation since the complex phase \com{in $\exp(-S_\mathrm{eff}^\mathrm{AF})$}
\com{appears} from $X_{N_{\tau}}$\com{via complex $m_x$}.
For low-momentum modes,
the phase cancellation mechanism exists and the weight cancellation is not
severe~\cite{Ohnishi:2012yb}.
An imaginary part in the modified mass term 
\com{involves} $\varepsilon_x$. 
In \com{the} case\com{ where} the $\pi$ fields 
take \com{a constant} 
amplitude\com{, the imaginary
contribution from one site in \equref{Eq:SeffAF} cancels with that in the
nearest neighbor site contribution}.
Then\com{ we} expect that \com{the weight cancellation is not severe} 
as long as low momentum modes contribute.
\com{By comparison, we may} have the severe weight cancellation when high momentum modes dominate. 
In Sec.~\ref{Sec:Results}, we quantitatively confirm these high momentum contributions.

\subsection{Chiral Angle Fixing}
\begin{figure}[tb]
 \begin{center}
  \includegraphics[width=105mm]{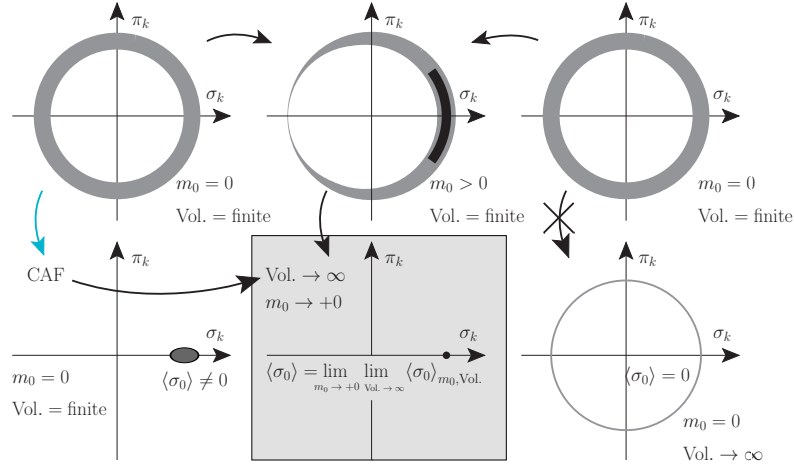}
 \end{center}
 \caption{Schematic picture of CAF method.
In order to obtain the value of chiral condensate,
we put a finite mass, first 
\com{take} thermodynamic limit 
and 
\com{finally take the chiral (massless)
limit} (like center bottom figure). 
In CAF, we take chiral rotation to make the $\pi_{0}$ field 
\com{disappears}.
Then, we get the finite chiral condensate (left bottom figure)\com{, which
would be close to the correct value}.
}
 \label{Fig:CAF}
\end{figure}

We develop a way to \com{define the chiral 
condensate and the chiral susceptibility on a finite volume lattice
in the chiral limit}.
Since the chiral transformation mixes the scalar mode, $\sigma_0$
\tcom{$=\sum_\tau \sigma_{\bm{k}=0,\tau}/N_\tau$,}
and its chiral partner, the pseudoscalar mode,
$\pi_0=\sum_\tau (-1)^\tau \pi_{\bm{k}=0,\tau}/N_\tau$, as

 \begin{align}
  \left( 
    \begin{array}{c}
      \sigma_0' \\
      \pi_0'
    \end{array}
  \right)= \left(
    \begin{array}{cc}
      \cos \alpha & \sin \alpha \\
      -\sin \alpha & \cos \alpha
    \end{array}
  \right)\left(
    \begin{array}{c}
      \sigma_0 \\
      \pi_0
    \end{array}
  \right),
\label{Eq:CAF}
\end{align}
the order parameter $\expv{\sigma_{0}}$ vanishes 
on \com{a} finite size lattice\com{ in the chiral limit}. 
\com{Rigorously, we need to take the thermodynamic limit 
followed by $m_0 \to 0$
as shown in \figref{Fig:CAF}, then we obtain finite $\expv{\sigma_0}$ 
and zero $\expv{\pi_0}$.
This procedure is time-comsuming, and}
the root mean square order parameter 
\com{is adopted} to avoid these problems\com{ in spin systems~\cite{MCsim1}}.
\com{We here} propose \com{a} similar
method, Chiral Angle Fixing (CAF).
In order to get the \tcom{appropriate} chiral condensate in the chiral limit,
we rotate \com{$\sigma_0$ and $\pi_0$}
\com{onto} the $\sigma_{0}$ axis.
Other auxiliary fields are also rotated by $\alpha$ as in \equref{Eq:CAF}
in each Monte-Carlo \com{configuration},
where $\alpha =\arctan (\pi_{0}/\sigma_{0})$. 
We use these new fields to \com{obtain} the order parameters\com{ 
and other quantities}.

We consider CAF from a different point of view.
Chiral condensate is described by
\begin{eqnarray}
\expv{\sigma_{0}} &=& 
\frac{1}{Z}\int \mathcal{D} \left[ \sigma_{0}, \pi_{0}\right]\ 
\prod_{(\bk ,\omega)\not=(\bm{0},0)}
%\prod_{\sigma_{\bk,\omega},\pi_{\bk,\omega}}
\mathcal{D}\tcom{\left[ \sigma_{\bk,\omega}, \pi_{\bk,\omega}\right]}\ \sigma_{0} \ 
e^{-S(\sigma_{\bk, \omega},\pi_{\bk,\omega},
\phi=\sqrt{\sigma_{0}^{2}+\pi_{0}^{2}},\alpha)}\ \nonumber \\
&=& 
\frac{1}{Z}\int \mathcal{D} \left[ \phi, \alpha\right]\ \phi \cos\alpha 
\prod_{(\bk ,\omega)\not=(\bm{0},0)}
\int \mathcal{D} \tcom{\left[\phi_{\bk,\omega}, \alpha_{\bk,\omega}'\right] }
\ e^{-S(\phi_{\bk, \omega},\alpha_{\bk,\omega}',\phi,0)} = 0
,
\label{Eq:bCAF}
\end{eqnarray}
where we define \com{$\sigma_{\bk,\omega},\ \pi_{\bk,\omega}$ 
as the Fourier transform of auxiliary fields}
%all modes 
other than $\sigma_0 ,\pi_0$ mode\com{s}.
$\phi_{\bk, \omega}$ and $\alpha_{\bk,\omega}$ are chiral \com{radius} 
and chiral angle in regard to each chiral partner, respectively.
\tcom{We also use a relation, 
$S(\phi_{\bk, \omega},\alpha_{\bk,\omega},\phi,\alpha)
=S(\phi_{\bk, \omega},\alpha_{\bk,\omega}'=\alpha_{\bk,\omega}-\alpha,\phi,0)$.}
According to \equref{Eq:bCAF}, chiral condensate is identically equal to zero. 
After we take CAF method,
we fix $\alpha$ to $\sigma_{0}$ axis and $\pi_0 =0$. We obtain finite chiral condensate
\begin{eqnarray}
\expv{\sigma_0 } 
= \frac{1}{Z} \int \mathcal{D} \sigma_0 \ \sigma_{0} 
%\prod_{\phi_{\bk ,\omega},\alpha_{\bk,\omega}}
\prod_{(\bk ,\omega)\not=(\bm{0},0)}
\int \mathcal{D} \tcom{\left[ \phi_{\bk,\omega},\alpha_{\bk,\omega} \right]} \ 
\ e^{-S(\phi_{\bk, \omega},\alpha_{\bk,\omega},\phi=\sigma_{0},0)}
\neq 0,
\label{Eq:aCAF}
\end{eqnarray}
in the Nambu-Goldstone (NG) phase.
We eventually obtain \com{chiral condensate} and chiral
susceptibility with finite peak. 
\com{Chiral} condensate \com{obtained} in CAF 
 should \com{mimic} the spontaneous chiral condensate
in the thermodynamic limit.

\section{Results of high momentum mode contributions} \label{Sec:Results}

We show the numerical results in this section, 
%and \tcom{assume} $T=\gamma^{2}/N_{\tau}$ \cite{Bilic92-2}.
where we have \tcom{assumed} $T=\gamma^{2}/N_{\tau}$ \cite{Bilic92-2}.
All results are carried out under CAF method, and error bars are estimated by the jack-knife method.
We quantitatively confirm high momentum auxiliary field mode contributions 
to the average phase factor $\expv{e^{\mrm{i}\delta}}$
and the order parameters.
We cut off high momentum modes with the 
%squared spatial kinetic momentum of auxiliary fields 
squared spatial momentum of auxiliary fields 
($\sum_{j=1}^{d} \sin^2 k_j$) 
more than the parameter $\Lambda$, 
and examine their effects. 
The parameter $\Lambda$ runs from 0 through $d=3$. When $\Lambda$ is equal to 3, we consider all Monte-Carlo configurations. 
By comparison, we only include the lowest momentum
modes when $\Lambda$ is equal to 0.
%%%%%%%%%%%%%%%%%%%%%%%%%%%%%%%%%%%%%%%%%%%%%%%%%%%%%%%%%%%%%%%%%%%%%%%%%%%%%%%%
%  figure
%%%%%%%%%%%%%%%%%%%%%%%%%%%%%%%%%%%%%%%%%%%%%%%%%%%%%%%%%%%%%%%%%%%%%%%%%%%%%%%%
\begin{figure}[tb]
%\PSfig{7.5cm}{L4N4.eps}~\PSfig{7.5cm}{PB6.eps}
%  \begin{minipage}{13cm}
    \begin{center}
\includegraphics[width=5.2cm]{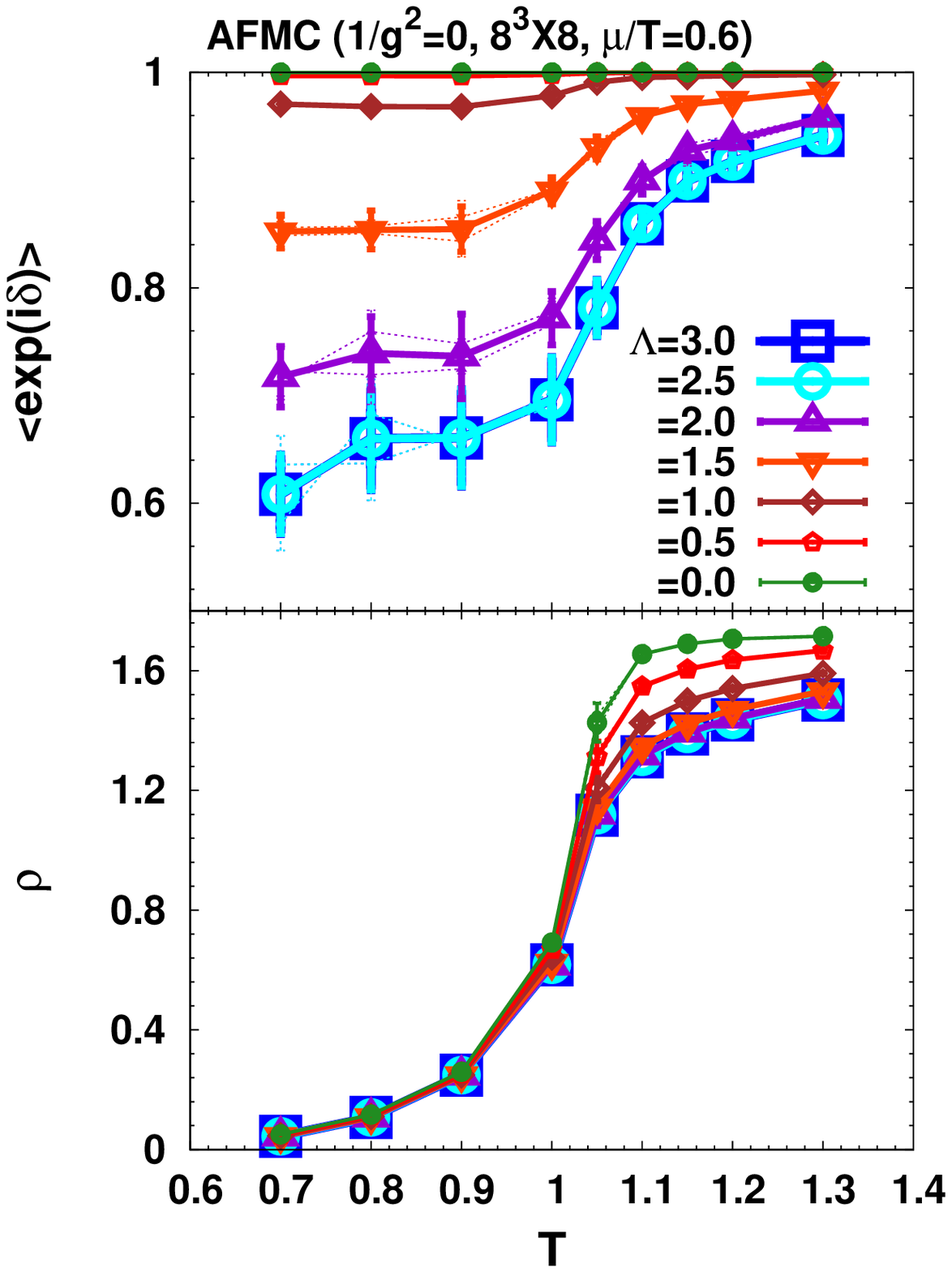}%
~\includegraphics[width=5.2cm]{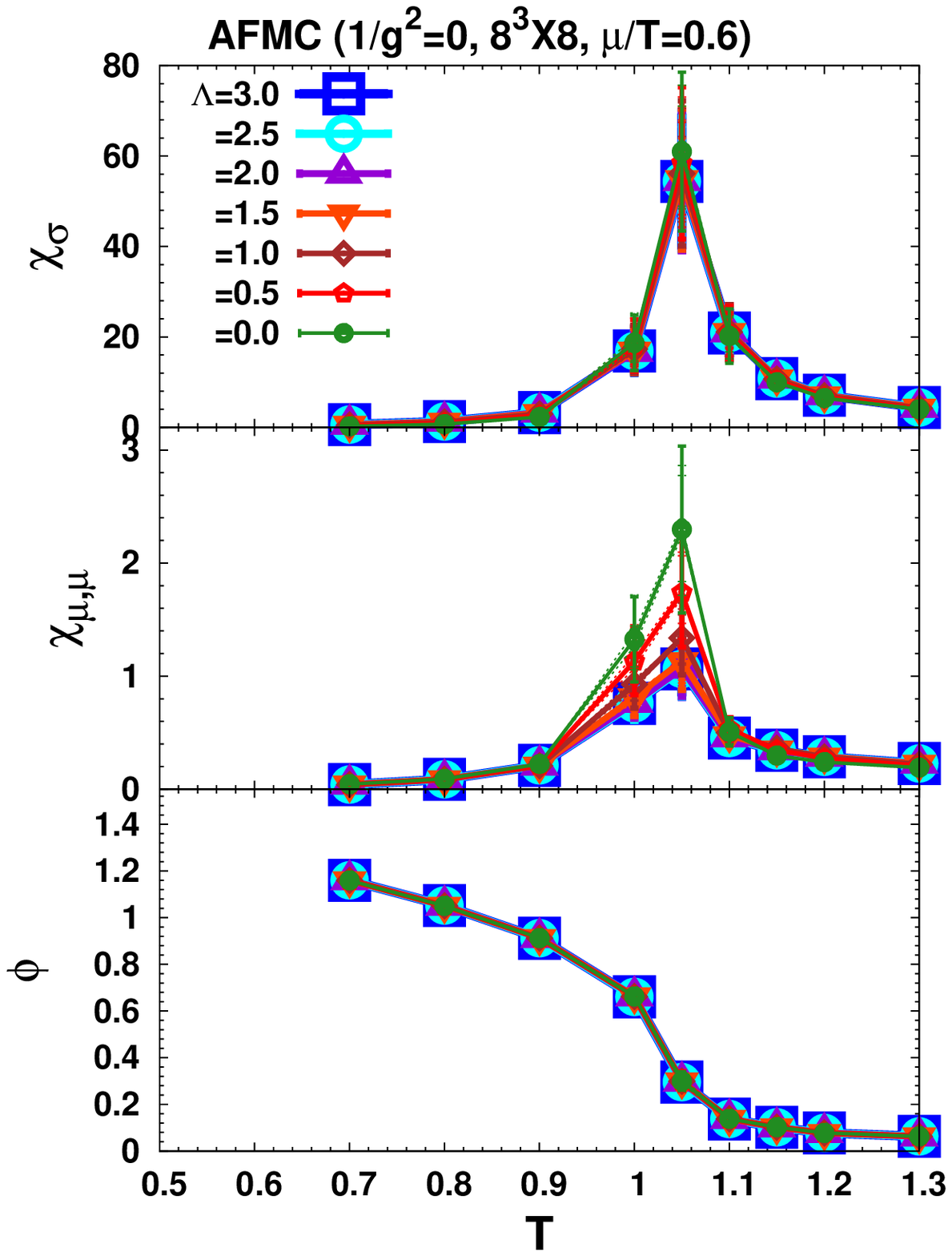}
    \end{center}
%  \end{minipage}
%~\hspace*{1cm}~
%  \begin{minipage}{6.5cm}
%    \begin{center}
%    \end{center}
%  \end{minipage}
\caption{Cut-off parameter $\Lambda$ dependence of average phase factor (left top), quark number density (left bottom), chiral susceptibility (right top), quark number susceptibility (right middle),
and chiral condensate (right bottom) as a function of temperature
on a $8^4$ lattice in the chiral limit for fixed $\mu /T=0.6$. 
}
\label{Fig:cut}
\end{figure}
%%%%%%%%%%%%%%%%%%%%%%%%%%%%%%%%%%%%%%%%%%%%%%%%%%%%%%%%%%%%%%%%%%%%%%%%%%%%%%%%
%  end figure
%%%%%%%%%%%%%%%%%%%%%%%%%%%%%%%%%%%%%%%%%%%%%%%%%%%%%%%%%%%%%%%%%%%%%%%%%%%%%%%%

As discussed in Sec. \ref{Sec:formalism}, 
the phase cancellation might be severe when high momentum modes contribute 
to the effective action.
We could expect that the weight cancellation becomes weaker 
if we cut off the high momentum modes. 
In \figref{Fig:cut}, we show the average phase
factor on a $8^4$ lattice for fixed $\mu /T =0.6$ in the chiral limit. 
We find that the weight cancellation weakens when $\Lambda$ goes to 0. 
This is consistent with our consideration of the phase cancellation with low momentum modes, 
and we could conclude that high momentum modes lead to \com{strong} weight cancellation.

We show the chiral condensate after CAF in the right bottom panel of \figref{Fig:cut}. 
The behavior of the chiral condensate does not depend on the cut-off parameter $\Lambda$. 
This is because integration variables in AFMC
are the mesonic auxiliary fields ($\sigma_{\bk,\tau}, \pi_{\bk,\tau}$), and the scalar and pseudoscalar mode\com{s} are comprised of the lowest modes of the integration variables. 
In \figref{Fig:cut}, we also plot the cut-off dependence of quark
number density, chiral susceptibility and quark number susceptibility. 
These results demonstrate \com{that} no
cut-off dependence as long as the parameter $\Lambda$ is larger than 2. 
Revisiting the average \com{phase} factor, we know the statistical
weight cancellation weakens in cases where $\Lambda$ is less than 2.5. 
Therefore, these results indicate
that there exists a optimal cut-off $\Lambda_o$, 
where the behavior of the order parameters does not change and
we could improve the reliability of numerical simulation. 
There is a possibility
to investigate the QCD phase diagram on a larger lattice by cutting off or approximating the high
momentum modes without changing the behavior of order parameters.

%%%%%%%%%%%%%%%%%%%%%%%%%%%%%%%%%%%%%%%%%%%%%%%%%%%%%%%%%%%%%%%%%%%%%%%%%%%%%%%%
%  figure
%%%%%%%%%%%%%%%%%%%%%%%%%%%%%%%%%%%%%%%%%%%%%%%%%%%%%%%%%%%%%%%%%%%%%%%%%%%%%%%%
%%%%%%%%%%%%%%%%%%%%%%%%%%%%%%%%%%%%%%%%%%%%%%%%%%%%%%%%%%%%%%%%%%%%%%%%%%%%%%%%
%  end figure
%%%%%%%%%%%%%%%%%%%%%%%%%%%%%%%%%%%%%%%%%%%%%%%%%%%%%%%%%%%%%%%%%%%%%%%%%%%%%%%%

Finally, we show the results of the chiral phase transition by using CAF. 
The order of the chiral phase transitions are deduced from the distribution of chiral condensate instead of considering finite size scaling of chiral susceptibility due to small lattice size\cite{Ohnishi:2012yb}. 
The distribution analysis indicates that the order of transition is the 1st (2nd or cross-over) order at high (low) $\mu$.
We deduce the would-be 1st order phase boundary as the point 
where the minimum of the effective potential changes discontinuously.
We obtain the would-be 2nd order phase boundary by fitting the peak of chiral susceptibility with the quadratic function. 
The error bars include both statistical and systematic errors for the (would-be) 2nd order transition. 
The \com{present} phase boundary is consistent with Monomer-Dimer-Polymer (MDP) simulation result \cite{Fromm,Unger}.
Both results show the NG phase is suppressed at low $\mu$ and broadened at high $\mu$ compared to the mean field (MF) results. 
In high $\mu$ region, the NG phase is enlarged as the temporal lattice size $N_{\tau}$ becomes large. This behavior is also seen in MDP, 
\com{which is also extrapolated to}
$N_{\tau} \rightarrow \infty$.

\section{Summary}
We have discussed the \com{mechanism for} the sign problem 
\com{to appear} in the auxiliary field Monte-Carlo (AFMC) method based on an effective action, $S_{\mrm{eff}}^{\mrm{AF}}$. %, 
The root cause  
of the sign problem in AFMC is extended Hubbard-Stratonovich (EHS) transformation, 
since we must introduce complex terms in bosonization.
The complex terms in the effective
action is eventually in the spatial diagonal parts of the fermion 
\com{matrix} (the modified mass term\com{s}).
The statistical weight cancellation arises when we numerically integrate out auxiliary fields on the basis of $S_{\mrm{eff}}^{\mrm{AF}}$. 
We quantitatively confirm 
that the high momentum modes contribute to the weight cancellation.
We also give a new approach 
\com{respecting} chiral symmetry
named as \com{the} Chiral Angle Fixing (CAF) method. 
We rotate all \com{auxiliary} field \com{modes}
in each Monte-Carlo \com{configuration} 
by \com{fixing the}
chiral angle, the angle between the scalar and the pseudoscalar 
modes, to zero. 
We acquire the finite chiral condensate and the chiral
susceptibility with a peak respecting chiral symmetry in CAF.
Finally, we find that the chiral phase transition boundary in CAF is consistent with another method called Monomer-Dimer-Polymer (MDP) simulations \cite{Fromm,Unger}. 

\section{Acknowledgement}
TI is supported by Grants-in-Aid for the Japan Society for Promotion of Science (JSPS) Research Fellows (No.~25-2059).
This work was supported in part
by Grants-in-Aid for Scientific Research
from the Japan Society for the Promotion of Science (JSPS)
(Nos.\
%  20540265, % TK
%  23105713, % HI (Innovative Areas)
  23340067, % TK (Kiban B)
  24340054, % A.Nakamura (Kiban B)
  24540271% AO (Kiban C)
%  10J03314 % T.Nakano JSPS
%  and innovative area
%  24105001, % AO (NS, X01)
%  24105008, % AO (NS, D01)
%for Scientific Research on Innovative Areas (No. 2004: 23105713).} 
),
by Grant-in-Aid for Innovative Areas
from the Ministry of Education, Culture, Sports, Science and Technology
of Japan (MEXT)
(Area No.~2404, Nos.~24105001, 24105008), % NS (X01 (Tamura), D01 (Ohnishi))
by the Yukawa International Program for Quark-Hadron Sciences.

%\bibliography{ref}
%\begin{thebibliography}{99}
%\bibitem{...} 
%....

%\end{thebibliography}

\end{document}